\documentclass[pdflatex,sn-mathphys-num]{sn-jnl}

\usepackage{graphicx}%
\usepackage{multirow}%
\usepackage{amsmath,amssymb,amsfonts}%
\usepackage{amsthm}%
\usepackage{mathrsfs}%
\usepackage[title]{appendix}%
\usepackage{xcolor}%
\usepackage{textcomp}%
\usepackage{manyfoot}%
\usepackage{booktabs}%
\usepackage{algorithm}%
\usepackage{algorithmicx}%
\usepackage{algpseudocode}%
\usepackage{listings}%

\graphicspath{{Figs/}}

\theoremstyle{thmstyleone}%
\theoremstyle{thmstyletwo}%

\theoremstyle{thmstylethree}%

\begin{document}

\title[Logistic Cellular Automata]{Emulating the logistic map with totalistic cellular automata}
\author[1,2,3]{\fnm{Franco} \sur{Bagnoli}}\email{franco.bagnoli@unifi.it}

\affil[1]{\orgdiv{Department of Physics and Astronomy}, \orgname{University of Florence and CSDC}, \orgaddress{\street{via G. Sansone, 1}, \city{Sesto Fiorentino}, \postcode{50019}, \state{(FI)}, \country{Italy}}}

\affil[2]{\orgname{INFN}, \orgdiv{Sect. Florence}}

\affil[3]{\orgdiv{Espace-Dev}, \orgname{University of Perpignan via Domitia}, \orgaddress{\street{52 Avenue Paul Alduy} \city{Perpignan}, \postcode{66000} \country{France}}}

\abstract{We investigate the conditions under which the mean-field formulation of a probabilistic, totalistic cellular automaton approximates the logistic equation. We show that this goal can be only fulfilled for an infinite-range neighborhood. We  numerically study the corresponding one-dimensional implementation, showing that the mean-field description is obviously approached by shuffling the configuration at each time step, but also by rewiring a fraction of links, either at each time step, or using the same random sampling once and for all, in the spirit of the ``small-world'' mechanism. We show that it is possible to obtain a good approximation of the logistic behavior already with a fraction of rewired links different from one. We also show that there is a bifurcation cascade of the density as a function of the fraction of the rewired links, and that this scenario also holds for a deterministic, totalistic CA with the same basic symmetries of the probabilistic one.}


\maketitle

\section{Introduction}

The logistic map~\cite{May1976} is an extremely popular subject of research: Google Scholar reports more than 2 millions entries searching with this term. 

This discrete-time dynamical system has been introduced by Robert May in 1976 (although a similar equation had been already studied by E. Lorenz in 1964~\cite{Lorenz1964}) as a model of the growth of a bounded population with discrete generations, as happens for insects that during the fall (autumn) lay eggs that will hatch the next spring. 

May's logistic map can be obtained by means of simple assumptions. 
When there is a  vanishing number of animals and great abundance of food, each insect will lay many eggs, and this causes an initial exponential growth of the population. However, when the number of insects becomes comparable to the maximal carrying capacity of the environment, the limited share of available food will imply a smaller average number of eggs. 

Let us denote by $x=x(t)$ the ratio of the number of animals at time $t$ with respect to the maximum population size. Neglecting spatial correlations, and assuming non-overlapping generations, the logistic map is 
\begin{equation}\label{eq:logistic}
     x' = a x (1-x),
\end{equation}
where $x'\equiv x(t+1)$. For vanishing values of $x$ this equation gives and exponential growth $x'=ax$, but when $x$ become comparable with one, the growth rate diminishes. Since $0\le x \le 1$, the control parameter $a$ is limited to the interval $0\le a \le 4$. 

\begin{figure}
    \centering
    \includegraphics[width=0.7\linewidth]{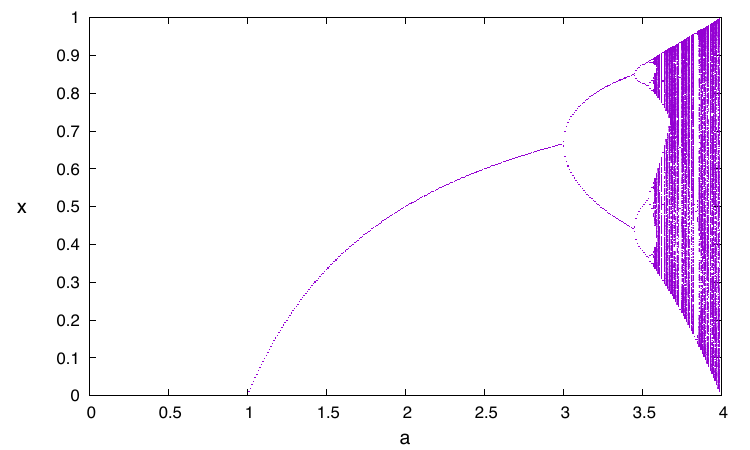}
    \caption{Bifurcation diagram of the logistic map, Eq.~\eqref{eq:logistic}. For each value of the parameter $a$ we plot $x(t)$ for 500 values of $t$, after a transient of 1000 time steps. }
    \label{fig:logistic}
\end{figure}

As noted already in the title of May's work~\cite{May1976}, by varying the parameter $a$ we get an interesting bifurcation diagram for the long-time values of $x$, as shown in Fig.~\ref{fig:logistic}, where we observe fixed points, limit cycles and chaotic oscillations.

The logistic map,  Eq.~\eqref{eq:logistic}, is a mean-field description which neglects spatial correlations. It is interesting to investigate which kind of microscopic model can give origin to such a macroscopic dynamics. A similar problem has been addressed in Ref.~\cite{Dzwinel2009}, and the conclusions are that some two-dimensional Boolean cellular automata, coupled with a random scattering of site values (random walks of individuals) and increasing neighborhood size, can effectively give origin to behavior of the average ``density'' similar to that of the logistic map. 

We can generalize this approach by investigating the conditions under which the mean-field description of a Boolean cellular automaton becomes similar or equal to the logistic map. Since there is no reason to prefer a particular ordering in the neighborhood of a cell, we restrict our investigation to the symmetric case, in which each neighbor contributes the same, i.e., totalistic cellular automata. In order to accommodate for a continuous control parameter $a$, we first investigate probabilistic automata, leaving the study of a deterministic one to the last part of this work. 

However, in a numerical implementation one needs to randomize the configuration in order to destroy the local correlations and approximate the mean-field description. This is particularly important for the values of parameters for which the mean-field description gives chaotic oscillations of the density. If the configuration is not homogenized, one rather obtains incoherent oscillations of local patches, which globally corresponds to noisy oscillations of the density around a fixed value, as already noticed by Boccara et al.~\cite{Boccara1994}. The homogenization goal can be achieved for instance using a shuffling procedure~\cite{Boccara1992}. 

However, the shuffling of a configuration is a process that is difficult to control, in the sense that it is not easy to impose a given degree of randomness. We can instead choose at random the neighbors of each cell, a procedure which is much easier to control, as illustrated in Section~\ref{sec:micro}. 

Moreover, we can profit of the ``small-world effect''. As noted by Watts and Strogatz~\cite{Watts1998}, by rewiring a fraction of links of an otherwise local model, one can get a smooth transition from the usual behavior of a system with local interactions, to the  mean-field one.
Indeed, as shown in models ranging from cellular automata~\cite{Bagnoli2013, Bagnoli2015} to the parallel Ising model~\cite{Bagnoli2016}, this rewiring procedure corresponds to different degrees of homogenization, revealed by bifurcation scenarios in models whose mean-field approximation is a chaotic map. 

The structure of this paper is the following. In Section~\ref{sec:model} we define precisely the totalistic, probabilistic cellular automaton. In Section~\ref{sec:mean-field} we shall obtain the conditions for which the mean-field description of such a model corresponds to the logistic map. 
In the same section we shall find that, as already noticed in Ref.~\cite{Dzwinel2009}, only for an infinite size of the neighborhood one can obtain the full range of the logistic map parameter. 
In Section~\ref{sec:micro} we shall present an implementation of a one-dimensional version of the previous cellular automaton, with different approaches to the mean-field behavior: shuffling the configuration, sampling at random the neighborhood at each time step, or rewiring a part of it only at beginning. We show that the resulting behaviors coincide, approximating that of the logistic map with finite-size noise. In the same section we shall study the the bifurcation diagram of probabilistic and deterministic cellular automata induced by different degrees of rewiring.

Finally, conclusions are drawn in the last section. 

\section{The cellular automaton model}
\label{sec:model}

We are interested in totalistic cellular automata of size $N$ and neighborhood size $R$. We denote by $s_i(t)\in \{0,1\}$ the value the cell $i$ at time $t$. We shall use the abbreviation $s_i\equiv s_i(t)$ and $s'_i\equiv s_i(t+1)$. The whole configuration will be denoted by $\boldsymbol{s}$.  

In order to define the problem, let us consider
the adjacency matrix 
\begin{equation}\label{eq:c}
 c_{ij} = \begin{cases} 1 &\text{if $s'_i$ depends on $s_j$, }\\
   0 &\text{otherwise.}
   \end{cases}
\end{equation}

The input connectivity of cell $i$ is given by $\sum_j c_{ij}$ and we impose that all cells have the same input connectivity $R$ (each row of $c_{ij}$ has exactly $R$ ones and $N-R$ zeros). The set of cells for which  $c_{ij}=1$ is called the neighborhood of cell $i$. Since in this section and the following one we are only interested in the mean-field description, we postpone the precise specification of the adjacency matrix to Section~\ref{sec:micro}.   

The totalistic character of the automaton corresponds to the fact that $s'_i$ is a function of the sum of the values of the cells in its neighborhood at time $t$. We shall denote this sum by $v_i$, 
\[
v_i = \sum_j c_{ij}s_j \qquad \text{with}\qquad 0 \le v_i \le R. 
\]

Let us define the transition probabilities $\tau(k)\equiv\tau(1|k)$, where $t(k)$ gives the probability that $s'_i=1$ if $v_i = k$. 
The evolution of the cellular automaton is given by the synchronous application of the same local rule 
\[
    s'_i = f(\tau(v_i)),
\]
where $f(x)$ takes value 1 with probability $x$ and zero otherwise. For a deterministic cellular automaton, $\tau(k)$ are either zero or one and  $f$ is the identity. 

We are interested in the density $x(t)$,
\begin{equation}\label{eq:dens}
 x(t) = \frac{1}{N} \sum_i s_i(t).
\end{equation}

The mean-field approximation consists in neglecting all spatial correlations, so that the evolution for the density $x$ becomes
\begin{equation}\label{eq:mf}
    x' = \sum_{k=0}^R \binom{R}{k} \tau(k) x^k (1-x)^{R-k},
\end{equation}
with $x'=x(t+1$), since for an uncorrelated configuration, the probability of having $k$ ones and $R-k$ zeros in the neighborhood of a cell is given by $x^k (1-x)^{R-k}$ times the number of possible permutations (the binomial coefficient).

\section{Determining the transition probabilities}\label{sec:mean-field}

Our goal is that of choosing the transition probabilities $\tau(k) \equiv \tau(1|k)$ so that Eq.~\eqref{eq:mf} coincides with the logistic one, Eq.~\eqref{eq:logistic},
with $0\le a\le a_M$, possibly with $a_M = 4$. 

Therefore, we want to compute $\tau(k)$ so that 
\begin{equation}\label{eq:correspondance}
\sum_{k=0}^R \binom{R}{k} \tau(k) x^k (1-x)^{R-k} = ax(1-x).
\end{equation}

We can obtain a series of relationships.
First of all, since the logistic map is invariant with respect to replacement of $x$ with $1-x$, we get 
\begin{equation}\label{eq:symmetry}
    \tau(k) = \tau(R-k).
\end{equation}

Setting $x=0$ we get $\tau(0)=0$ and, therefore, also $\tau(R)=0$. 
We can obtain another relation by setting $x=1/2$:
\begin{equation}\label{eq:sum}
    \frac{1}{2^R} \sum_{k=1}^{R-1} \binom{R}{k} \tau(k)  = \frac{a}{4}.
\end{equation}

We now proceed by equating the coefficients of the powers of $x$. Let us denote them by $b(k)$,
\[
\sum_{k=1}^{R-1} \binom{R}{k} \tau(k) x^k (1-x)^{R-k} = \sum_{k=1}^{R-1} b(k) x^k,
\]
in order to fulfill Eq.~\eqref{eq:correspondance}, we need to have
\[
\begin{cases}
    b(1) = a,\\
    b(2) = -a,\\
    b(k) = 0 & \text{for $k\ge 3$}.
\end{cases}
\]

By expanding the binomial coefficients we get, for the coefficient $b(1)$,
\[
    b(1) = R \tau(1) = a 
\]
and, therefore,
\[
    \tau(1) = \frac{a}{R}.
\]

The  coefficient of $x^2$ comes from  
\[
    \binom{R}{1}\tau(1) x(1-x)^{R-1} + \binom{R}{2}\tau(2) x^2 (1-x)^{R-2}.
\]
By taking the second term from the expansion of the coefficients of $\tau(1)$ and the first from that of $\tau(2)$, we get
\[
    -R[R-1] \tau(1) + \frac{R(R-1)}{2} \tau(2) = -a,
\]
where we grouped with square brackets the contribution coming from the expansion of $(1-x)^{R-k}$. Therefore, 
\[
    \tau(2) = \frac{2(R-2)}{R(R-1)} a .
\]

Similarly, for the coefficient  $b(3)$,
\[
    R\left[\frac{(R-1)(R-2)}{2}\right] \tau(1) - \frac{R(R-1)}{2} [R-2] \tau(2) + \frac{R(R-1)(R-2)}{3!} \tau(3) = 0;
\]
and replacing the values of the computed $\tau$, we get 
\[
    \frac{(R-2)}{2}\left((R-1) - 2(R-2) \right)a  + \frac{R(R-1)(R-2)}{3!} \tau(3) = 0.
\]

After simplifying, the final expression is
\[
    \tau(3) = \frac{3(R-3)}{R(R-1)} a.
\]

The procedure is similar for the other coefficients, and after some algebra, we obtain
\[
    \tau(k) = \frac{k(R-k)}{R(R-1)} a.
\]
 Notice that this expression reflects the symmetry of Eq.~\eqref{eq:symmetry}. 

We can check that the transition probabilities also satisfy Eq.~\eqref{eq:sum}.
By considering the expansion of the binomial,
\[
    \sum_{k=0}^R \binom{R}{k} x^k y^{R-k} = (x+y)^R,
\]
we see that 
\[
    xy \frac{\partial^2}{\partial x \partial y} \sum_{k=0}^R \binom{R}{k} x^k y^{R-k} = 
    \sum_{k=0}^R 
    \binom{R}{k} k(R-k) 
    x^k y^{R-k} = R (R-1) x y (x+y)^{R-2},
\]
so that, setting $x=y=1/2$ and rearranging a bit, we get Eq.~\eqref{eq:sum}:
\[
     \frac{1}{2^R} \sum_{k=1}^{R-1} \binom{R}{k} \tau(k)  =  \frac{1}{2^R} \sum_{k=0}^{R} \binom{R}{k} \frac{k(R-k)}{R(R-1)}a  = \frac{a}{4}.
\]

The problem of approximating the logistic equation is now that of determining $R$ such that the coefficients $\tau(k)$ are probabilities, i.e., between $0$ and $1$, which also establishes a maximum value $a_M(R)$. 

For a given value of $R$, the expression
\[
    \tau(k)=\frac{k(R-k)}{R(R-1)}a
\]
is a discretized parabola, with a maximum for $k=R/2$ (we can consider $R$ even for simplicity). Equating it to one, we get
\[
    a_M(R) = 4\frac{R-1}{R},
\]
so that we cannot really have $a_M=4$ for finite values of the neighborhood size $R$, i.e., $a_M(R)<a_M(\infty)=4$.

\section{One-dimensional cellular automata bifurcation diagrams}~\label{sec:micro}

From now on, we shall consider one-dimensional automata, even if similar reasoning can be applied to higher dimensions with small modifications.

In the case of a uniform lattice, the adjacency matrix $c$ of Eq.~\eqref{eq:c} is a circulant one, i.e., row $i+1$ is obtained  circularly shifting row $i$ by one position. 

\begin{figure}
    \centering
    \begin{tabular}{ccc}
    (a) & (b) & (c)\\
    \includegraphics[width=0.3\linewidth]{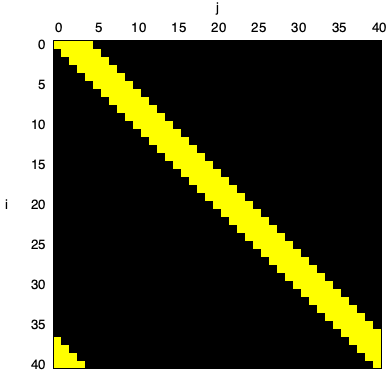} & 
    \includegraphics[width=0.3\linewidth]{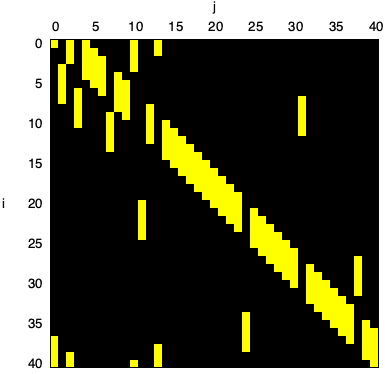} &
    \includegraphics[width=0.3\linewidth]{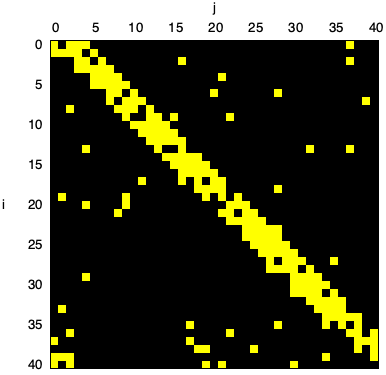}
    \end{tabular}
    \caption{Adjacency matrix for $N=41$, $R=5$, yellow dots marks connected neighbors. (a) Regular network (skewed neighborhood); (b) shuffled $p\simeq 0.2$; (c) rewired $p\simeq 0.2$.}\label{fig:adjacency}
\end{figure}

Notice that for quantities computed on the configuration at the same time, like the density $x$ of Eq.~\eqref{eq:dens},
it is irrelevant if we circularly shift all the rows of the matrix $c$ of the same amount. As an example, let us assume that cell $i$ depends on cells $i-1$, $i$, and $i+1$. Quantities computed at the same time take the same value if we instead make cell $i$ depend on cells $i$, $i+1$, $i+2$ or any other set of three consecutive cells, provided that the same shift applies to all cells. This consideration allows to define ``skewed'' neighborhoods, with all cells at the right or at the left of that under investigation, with the advantage of allowing the same notation for neighborhoods with an even or odd number of cells. 

For a cellular automaton with local interactions, the matrix $c$ is a band matrix. For row $i$ we can take the lower bandwidth to be $i$ and the upper bandwidth to be $i+R-1$, with periodic boundary conditions, see Fig.~\ref{fig:adjacency}-a for $N=41$ and $R=5$. Let us denote this matrix as $c^{(r)}_{ij}$.

Since we are interested in approximating the mean-field behavior, we need to homogenize the configuration. 
There are at least two ways of actually destroying correlations in a configuration, keeping the same density $x$: either shuffling it or rewiring incoming links. 

Let us examine the shuffling case. The basic procedure is the swapping of the values of two cells, say with indices $j$ and $k$, at time $t$. Cells that at time $t+1$ depended on  cell $j$  now depend on cell $k$ and vice versa. This is equivalent to swapping columns $j$ and $k$ in the matrix $c$, see Fig.~\ref{fig:adjacency}-b. 

In the rewiring case, we simply swap a number of entries in some rows of $c$, see Fig.~\ref{fig:adjacency}-c. 

In both cases, the input connectivity (the sum of entries in each row) remains the same. 

This procedure allows to quantify the degree $p$ of randomization of the matrix $c$: it is given by the number of entries that are different from the regular case, divided by $N$ and $R$:
\[
    p = \frac{1}{NR}\sum_{ij} |c^{(r)}_{ij} - c_{ij}|.
\]

\begin{figure}
    \centering
    \begin{tabular}{cc}
             (a) & (b) \\    
        \includegraphics[width=0.45\linewidth]{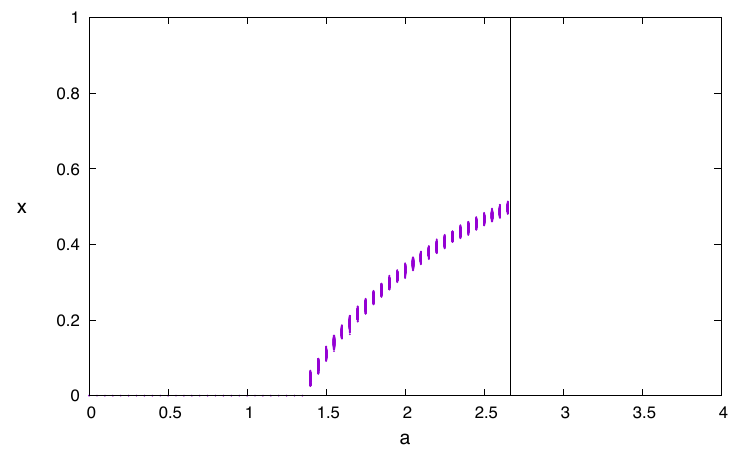} &
         \includegraphics[width=0.45\linewidth]{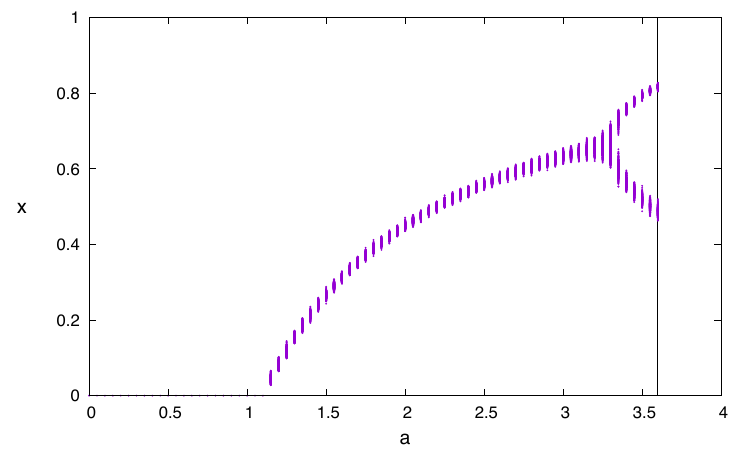}\\
             (c) & (d) \\    
        \includegraphics[width=0.45\linewidth]{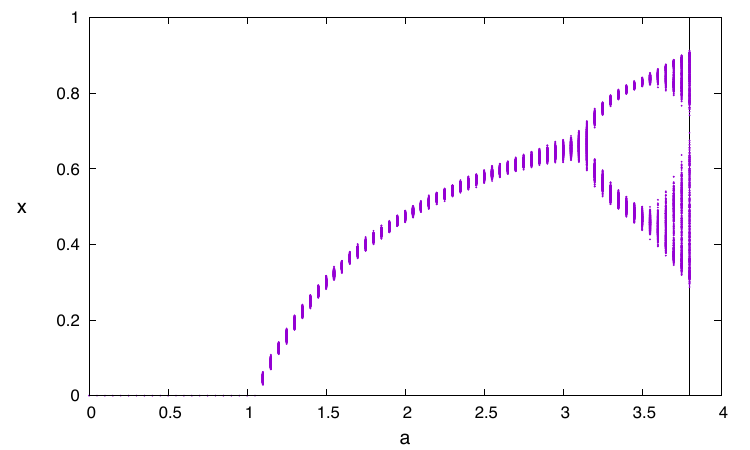} &
         \includegraphics[width=0.45\linewidth]{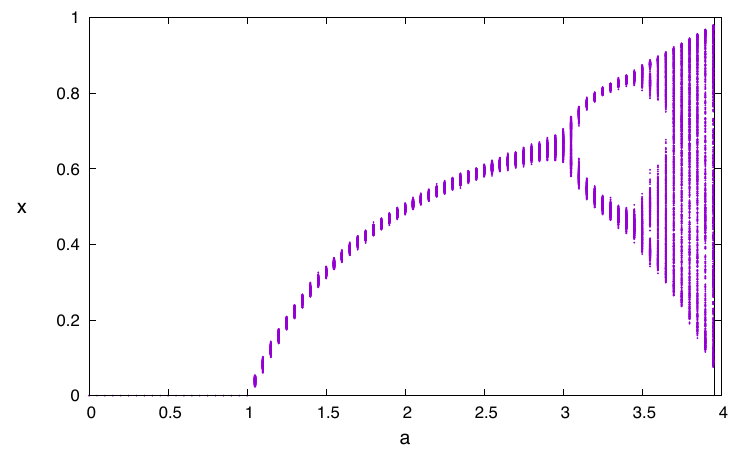}\\
    \end{tabular}
    \caption{Bifurcation diagrams of $x$  for $0\le a\le a_M$, $T=100$ steps and lattice size $N=10^4$, after a transient $S=10^3$ steps. The vertical line marks the value of $a_M$. The plots are the same for the shuffling, annealed and quenched procedure with $p=1$. (a) $R=3$; (b) $R=10$; (c) $R=20$ and (c) $R=100$.}
    \label{fig:bifurcCA}
\end{figure}

In the rewiring case, it is easy to generate matrices with a desired value of $p$, since one can simply assign a link to its regular position with probability $1-p$ or to a random position outside the band with probability $p$. For the shuffling case, one can swap columns as far as the resulting value of $p$ is less than the desired one (this procedure can take a long time for $p\simeq 1$). 

The above described procedures can be performed at each time step (regenerating the matrix $c$), or just once at the beginning (keeping the same $c$). We shall refer to the first case with the term ``annealed'' and to the second with the term ``quenched''. As we shall see, the effects of the quenched and the annealed procedures are  the same for large lattices, which essentially constitutes the small-world effect~\cite{Watts1998}. 

Let us fist consider the case $p=1$. As shown in Fig.~\ref{fig:bifurcCA}, the plot of the average density $x$ starts resembling the bifurcation diagram of the logistic map as $R$ increases, with a noise due to the finite size $N$ of the lattice. Numerically, we get the same results for the shuffling and the rewiring mechanisms, quenched or annealed.

\begin{figure}
    \centering
\includegraphics[width=0.7\linewidth]{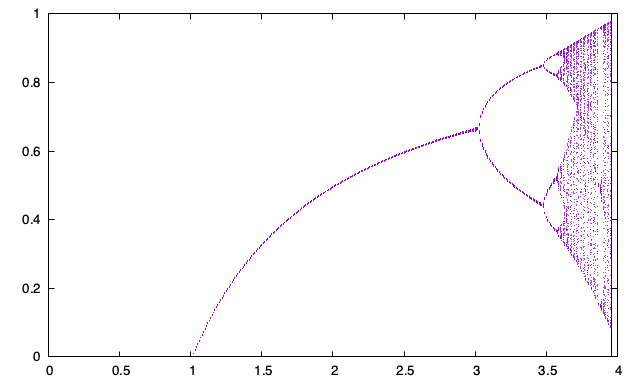}
    \caption{Bifurcation diagrams of the density $x$  for $0\le a\le a_M=0.96$,  $R=100$ and quenched rewiring, $T=100$ steps and lattice size $N=10^6$, after a transient $S=10^3$ steps. The vertical line marks the value of $a_M$. }
    \label{fig:bifurcCA1}
\end{figure}

Actually, the bifurcation diagram of the disordered cellular automaton approaches more and more that of the logistic map by increasing the lattice size. In Fig.~\ref{fig:bifurcCA1} the same diagram of Fig.~\ref{fig:bifurcCA}-d is shown, for $N=10^6$.

It is now interesting to examine how the small-world procedure (quenched rewiring the incoming neighborhood links with probability  $p$) modulates the approach to the logistic equation. 

\begin{figure}
    \centering
    \begin{tabular}{cc}
             (a) & (b) \\    
        \includegraphics[width=0.45\linewidth]{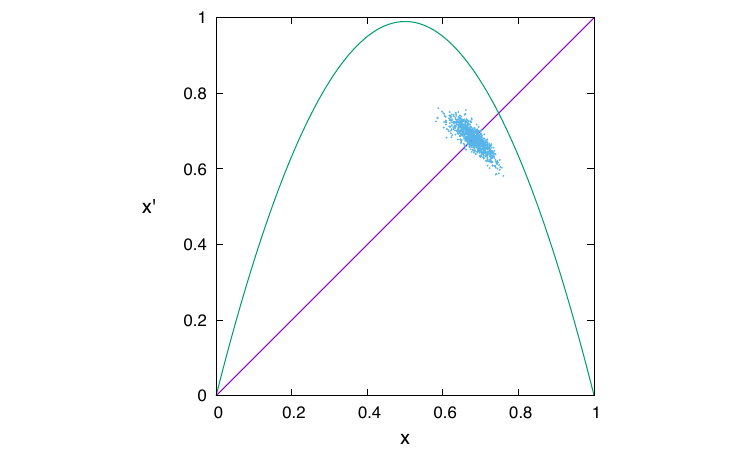} &
         \includegraphics[width=0.45\linewidth]{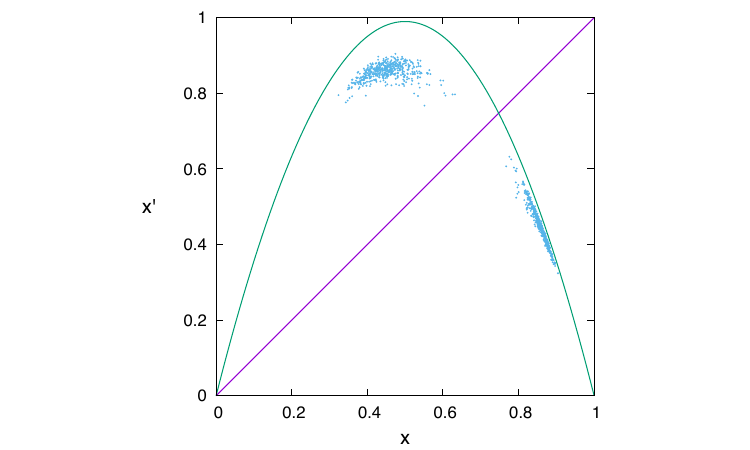}\\
             (c) & (d) \\    
        \includegraphics[width=0.45\linewidth]{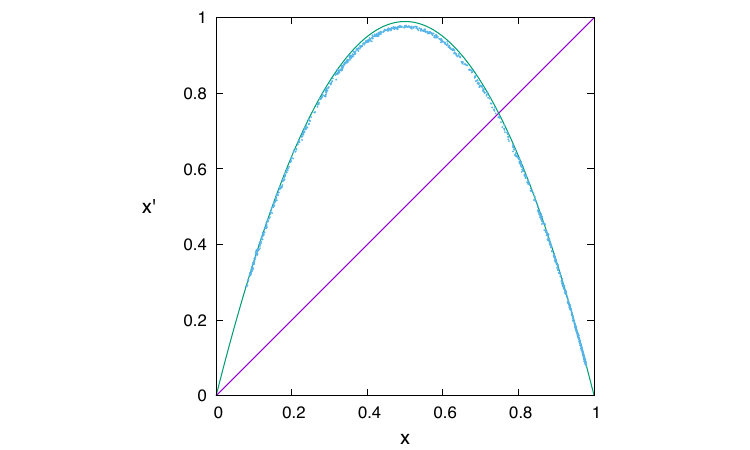} &
         \includegraphics[width=0.45\linewidth]{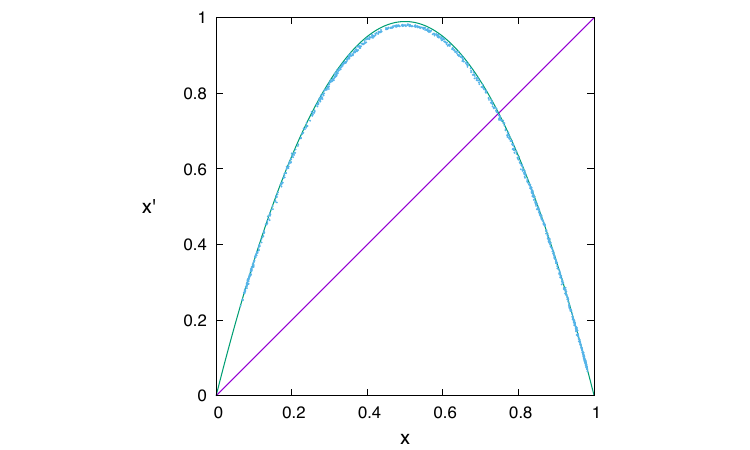}\\
    \end{tabular}
    \caption{Plot of the return map $x'\equiv x(t+1)$ vs $x\equiv(t)$ for $R=100$, $T=1000$ steps and lattice size $N=10000$, after a transient $S = 100000$ steps for various degrees of rewiring $p$ (dots) compared to the logistic map function (green line) for $a=a_M(R)=3.96$.  (a) $p=0.1$; (b) $p=0.2$; (c) $p=0.5$ and (d) $p=1.0$.}
    \label{fig:p}
\end{figure}

One can see from Fig.~\ref{fig:p} that for small values of $p$ the density is essentially constant, even for the maximum value $a=a_M$. Indeed, for a local neighborhood, the system is practically decoupled into almost independent domains, which oscillate in an incoherent way. 
By increasing $p$ the return map approaches the logistic curve. 

\begin{figure}
    \centering
    \begin{tabular}{cc}
    (a) & (b) \\
    \includegraphics[width=0.45\linewidth]{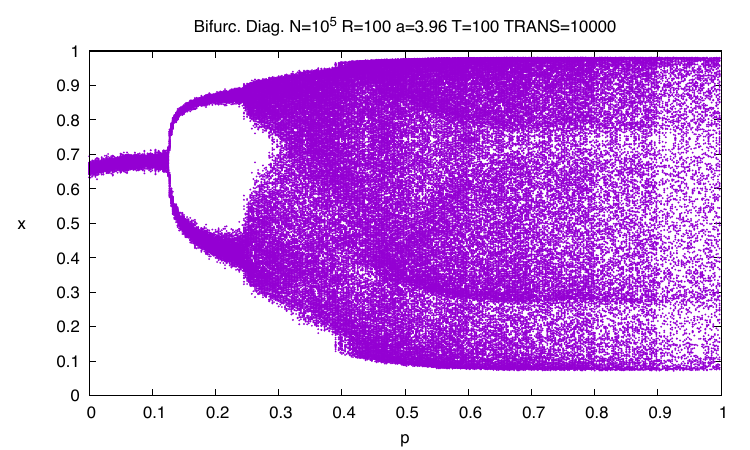} & 
    \includegraphics[width=0.45\linewidth]{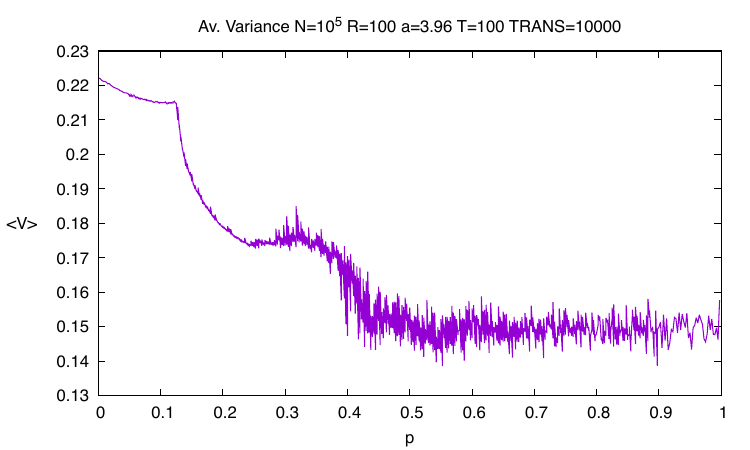}
    \end{tabular}
    \caption{(a) The bifurcation diagram of the density $x$ vs the rewiring probability $p$. (b) The average variance of the configuration $\langle V\rangle=(1/T)\sum_t V(\boldsymbol{s}(t))=(1/T) \sum_t x(t)(1-x(t))$ vs rewiring probability $p$. Parameters: $R=100$, $a=a_M=3.96$, $T=100$ steps and lattice size $N=10^5$, after a transient $S = 10^4$ steps.  }
    \label{fig:bifurcp}
\end{figure}

By looking at Fig.~\ref{fig:bifurcp}-(a), it is evident that the range of the density $x$ already approximates the final one for $p\gtrsim 0.6$ (even if there are other structures appearing for larger values of $p$). 

\begin{figure}
    \centering
    \begin{tabular}{cc}
    (a) & (b) \\
    \includegraphics[width=0.45\linewidth]{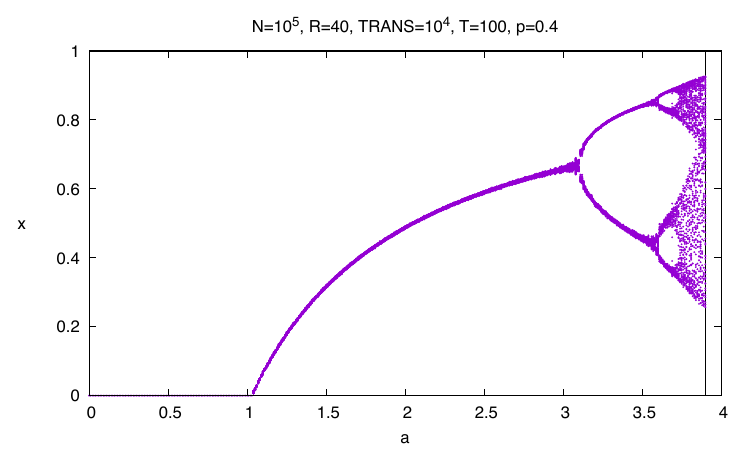} & 
    \includegraphics[width=0.45\linewidth]{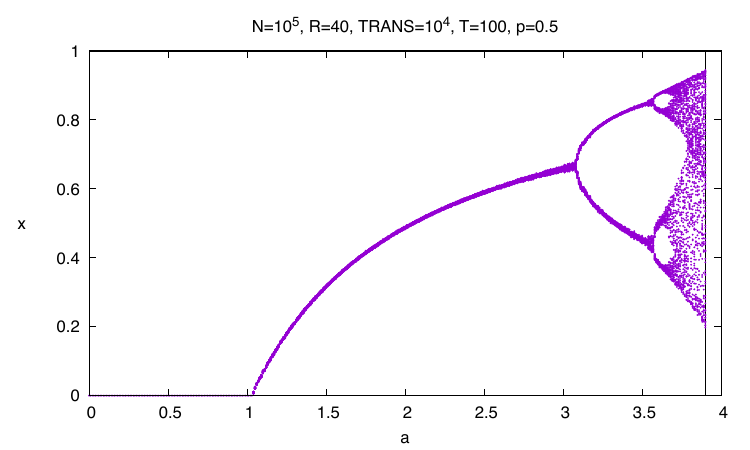} \\
    (c) & (d)\\
    \includegraphics[width=0.45\linewidth]{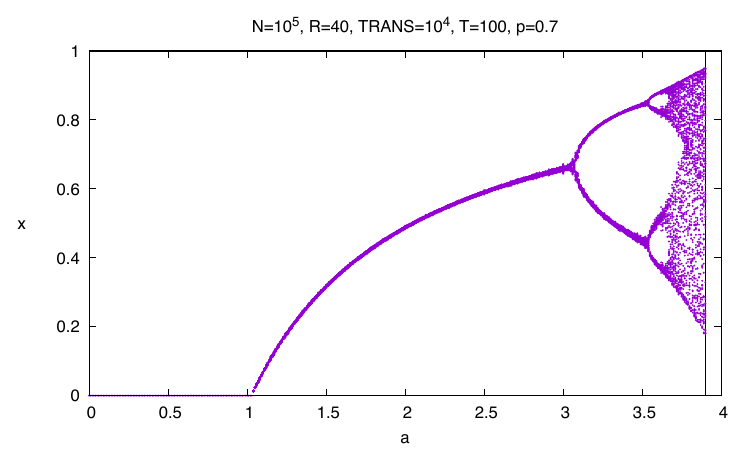} & \includegraphics[width=0.45\linewidth]{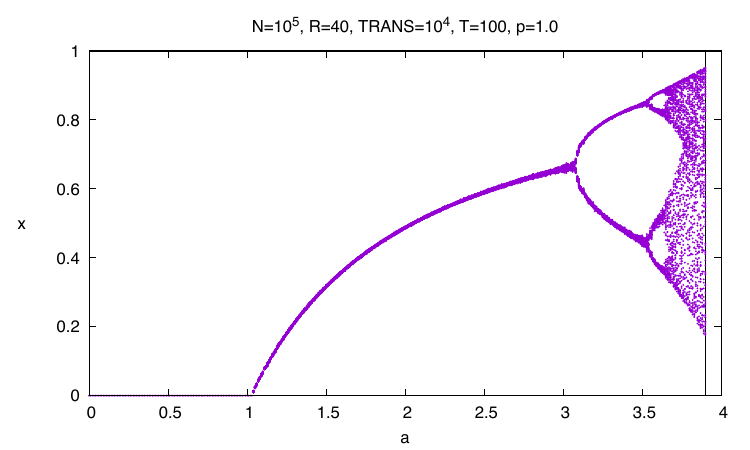}
    \end{tabular}
    \caption{The bifurcation diagram of the density $x$ vs the parameter  $a$ for various values of $p$ and $N=10^5$, $R=40$, $T=100$ after a transient $S=10^4$ steps. (a) $p=0.4$; (b) $p=0.5$; (c) $p=0.7$; (d) $p=1.0$. The vertical line marks the value of $a_M=3.9$.}
    \label{fig:bifurcp1}
\end{figure}

One can directly check the bifurcation diagrams for various values of the rewiring parameter $p$, as shown in Fig.~\ref{fig:bifurcp1}. One can see that near $a=a_M$ the location of the main bifurcations are almost already established for $p\gtrsim 0.6$, while, for $a\ll a_M$, the agreement is excellent even for smaller values of $p$. Clearly,  more details emerge by increasing the lattice size, as shown in Fig.~\ref{fig:bifurcCA1}.

\begin{figure}
    \centering
    \begin{tabular}{cc}  
    (a) & (b) \\
    \includegraphics[width=0.45\linewidth]{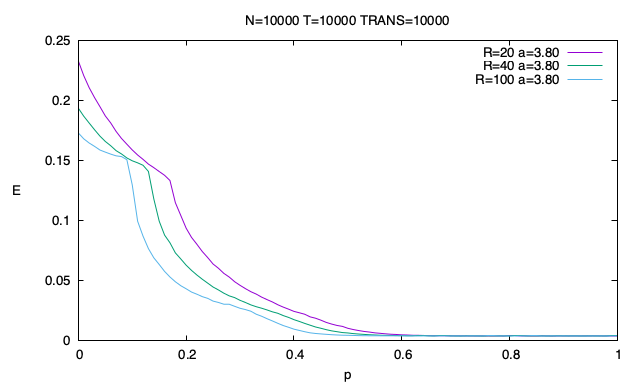} &
    \includegraphics[width=0.45\linewidth]{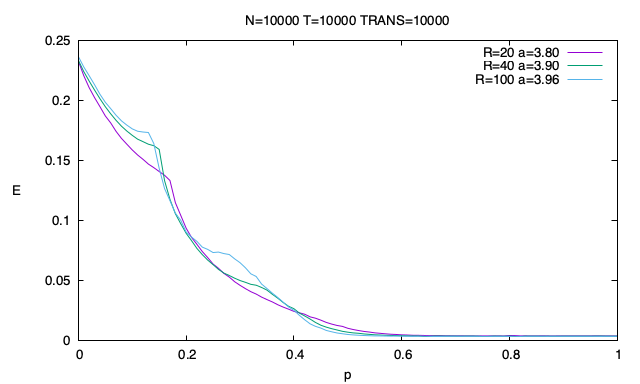}
    \end{tabular}
    \caption{Mean squared distance $E$ between the numerical return value $x'_\text{num}=x(t+1)$ and the one given by Eq.~\eqref{eq:logistic}, $x'_\text{theor}=ax(t)(1-x(t))$ vs the rewiring parameter $p$ for $R=20,40,100$, $N=1000$, $T=10000$, $\tau=10000$. (a) $a(20)=a(40)=a(100)=a_M(20)=3.8$, (b) $a(20)=a_M(20)=3.8$, $a(40)=a_M(40)=3.9$, $a(100)=a_M(100)=3.96$. The cusps corresponds to the bifurcation points, as shown in Fig.~\ref{fig:bifurcp} for $R=100$.}
    \label{fig:errorp}
\end{figure}

Finally, one can quantitatively compute the error $E$ by measuring the difference between the value of the density $x'_\text{num}=x(t+1)$ and the value expected by the logistic map of Eq.~\eqref{eq:logistic}, $x'_\text{theor}=ax(t)(1-x(t))$:
\begin{equation}\label{eq:E}
    E = \sqrt{\frac{1}{T} \sum_{t=S}^{S+T} \left[x(t+1) - ax(t)(1-x(t))\right]^2},
\end{equation}
where $S$ is the transient. The error $E$ as a function of the rewiring probability $p$ for the quenched case is reported in Fig.~\ref{fig:errorp}, for three values of $R$ and two cases: keeping the parameter $a$ fixed to the maximum value for $R=20$, $a=a_M(20)=3.8$  (Fig.~\ref{fig:errorp}-a), and for $a$ equal to the maximim values $a_M(R)$ for different values of $R$ (Fig.~\ref{fig:errorp}-b). In both cases one can see that the error becomes quite small already for $p\simeq 0.6$. 

The small-world transition of Figs.~\ref{fig:p} and \ref{fig:bifurcp} can be interpreted as a synchronization process: the density $x$ can oscillate in a quasi-deterministic way instead of fluctuating around an average value only if the sites in the configuration are coherent (have similar values for many neighborhoods). However, since each site value can only assume values $0$ and $1$, intermediate values of $x$ are possible only if large portions of the configuration take  different values. So the maximum coherence is for $x$ near zero or one. 

For Boolean variables, the variance $V(\boldsymbol{s})$ of configuration $\boldsymbol{s}$ is 
\[
V(\boldsymbol{s}) = \frac{1}{N^2}\left(N\sum_i s_i^2 - \Bigl(\sum_i s_i\Bigr)^2\right)=x(\boldsymbol{s})(1-x(\boldsymbol{s})),
\]
where $x$ is the density defined in Eq.~\eqref{eq:dens}, since $s_i^2 = s_i$. This implies that the smallest values of the variance correspond to the extreme values of the density $x$. In the numerical simulations, the variance is averaged over time for a certain number of steps. 

Another way of looking at the bifurcation diagram of Fig.~\ref{fig:bifurcp} is that of increasing synchronization. Indeed, the variance of the configuration averaged over time,  Fig.~\ref{fig:bifurcp}-(b), shows a steep decrease in correspondence of the first bifurcation (appearance of a limit cycle) for $p\simeq 0.2$, and stabilizes at its minimum value for $p\gtrsim 0.6$, where indeed the distribution approaches the final one, see Fig.~\ref{fig:p}-(c) and (d). 

\begin{figure}
    \centering
    \includegraphics[width=0.7\linewidth]{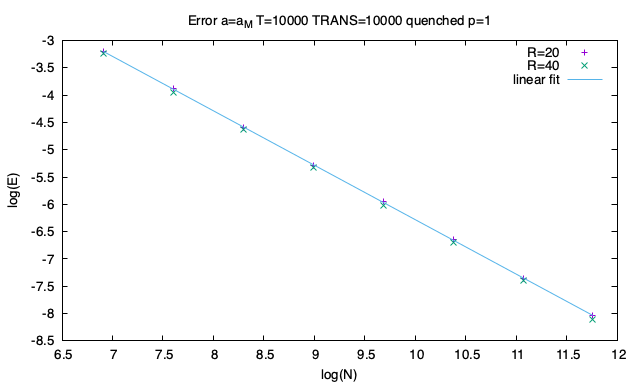}
    \caption{Mean squared distance $E$, Eq.~\eqref{eq:E}, vs the lattice size $N$ ($100\le N\le 128000$) for $R=20,40$, $a=a_M(R)$, $p=1$. The line is a linear fit in the log-log plot, with a slope $-0.975\pm 0.02$, consistent with the standard scaling due to uncorrelated noise. }
    \label{fig:errorN}
\end{figure}

The scaling of the error $E$ of Eq.~\eqref{eq:E} with the lattice size $N$ for full rewiring $p=1$ is reported in Fig.~\ref{fig:errorN}. The slope of the fit line in the log-log plot is $-0.975\pm 0.02$, showing that the error behaves like a standard sampling noise. 

The bifurcation diagram induced by randomization can be appreciated also for deterministic cellular automaton, which do not have any continuous adjustable parameters. This investigation has been done by Boccara and Roger for some totalistic cellular automata using a shuffling procedure~\cite{Boccara1992}. 

In particular, we investigate the totalistic case $R=7$, with transition ``probabilities'' (actually deterministic) such that 
\[
 s'_i = \begin{cases} 
   0 & \text{if $k = 0$ or $k=7$ },\\
   1 & \text{otherwise}.
   \end{cases}
\]
Since the value of $s'_i$ for $k=7,6,\dots,0$ is $0,1,1,1,1,1,1,0$, i.e., 126 in the base-two representation, this rule can be identified by this number. 
Notice that this deterministic automata obeys the  requirements of Eq.~\eqref{eq:symmetry} and $\tau(0)=\tau(R)=0$. 

The relative mean-field equation is
\[
 x' = x(1-x)[7(1-2x+3x^2-2x^3+x^4)]
\]
and one can see that this map has the structure of the logistic one, with a quadratic maximum for $x=1/2$. It is not surprising that the density of the related microscopic model exhibits a  bifurcation diagram similar to that of the logistic map when approaching the mean-field behavior~\cite{Boccara1992}, as suggested by the universality of chaos mechanisms~\cite{Feigenbaum1983}. 

\begin{figure}
    \centering
    \begin{tabular}{cc}
    (a) & (b) \\
\includegraphics[width=0.45\linewidth]{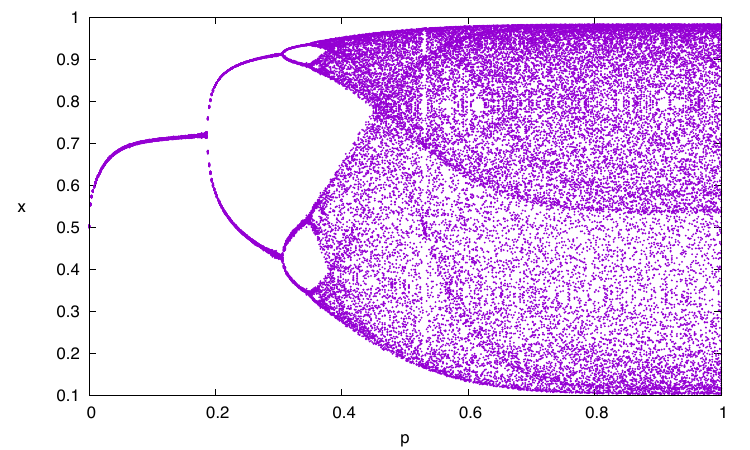} &
\includegraphics[width=0.45\linewidth]{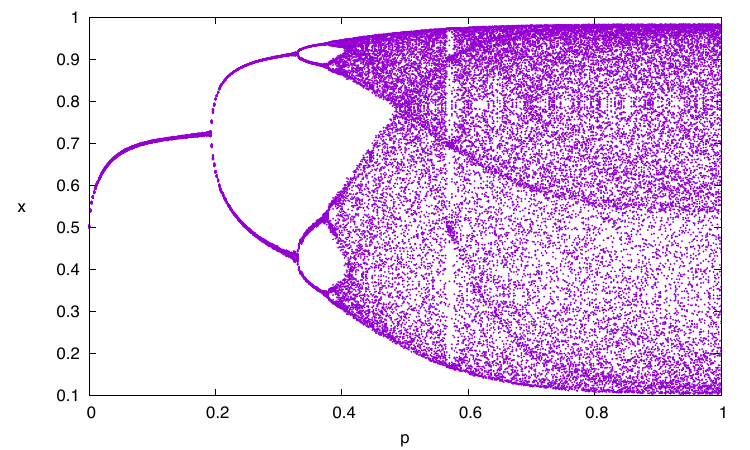}
\end{tabular}
        \caption{Bifurcation diagram of the $R=7$ totalistic cellular automata rule 126. $T=100$ steps of the density $x$ as a function of the rewiring fraction $p$ for $N=10^6$.  (a) Annealed version, links are rewired at each time steps, transient  $S=10^3$ steps; (b) quenched version, links are rewired only at beginning and kept fixed, transient $S=2\cdot 10^3$ steps.}
        \label{fig:126}
\end{figure}

This model is also a good testbed for the small-world hypothesis. Indeed, one can see in Fig.~\ref{fig:126} that the bifurcation diagram of the rule 126 as a function of the disorder $p$ shows several period-doubling and intermittent bifurcations, and that quenched and annealed rewiring give similar results, with small differences in the location of the bifurcation points. The annealed approximation is much more demanding in computational terms. 

As for the probabilistic case, for the regular lattice the dynamics is ``chaotic'' at the microscopic level, so that distant patches are uncorrelated and the average density is about $1/2$. The shuffling or rewiring of lattice behaves as a spatial homogenization mechanism, so that the density of the configuration starts oscillating following the mean-field equation and one has a transition between the microscopic to the macroscopic chaos. 

\begin{figure}
    \centering
    \begin{tabular}{ccc}
    (a) & (b) & (c) \\
    \includegraphics[width=0.3\linewidth]{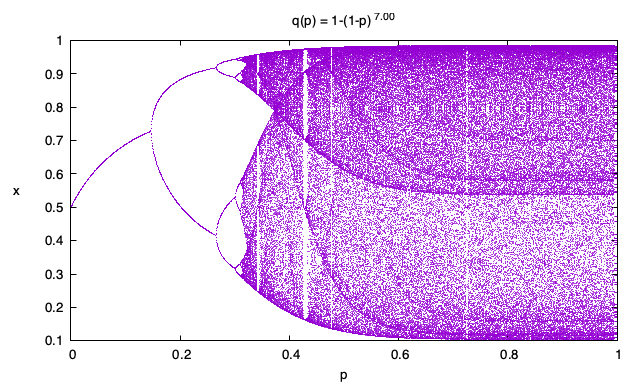} & 
    \includegraphics[width=0.3\linewidth]{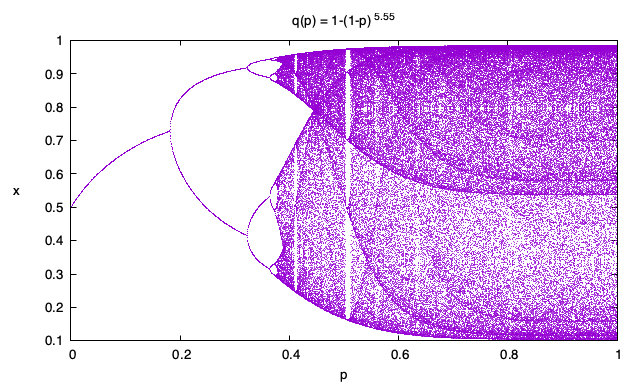} & 
    \includegraphics[width=0.3\linewidth]{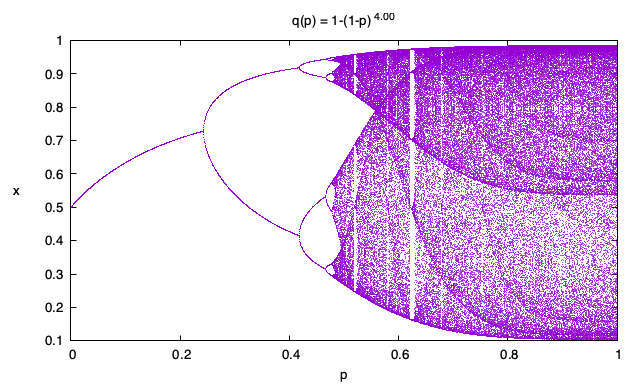}
    \end{tabular}
    \caption{The bifurcation diagram of the map of Eq.~\eqref{eq:126mf} for $q(p)=1-(1-p)^y$, with (a) $y=7$; (b) $y=5.55$; (c) $y=4$. Plot of 500 values of $x(t)$ after a transient of 500 time steps.}
    \label{fig:126mf}
\end{figure}

One can try to approximate this transition by assuming that the dynamics of the density is given by a combination of its ``chaotic'' value, $x=1/2$, and that of the mean field, i.e., 
\begin{equation}\label{eq:126mf}
    x' = (1-q(p))\frac{1}{2} + q(p) x(1-x)[7(1-2x+3x^2-2x^3+x^4)],
\end{equation}
with $q(p)$ being an appropriate function of $p$. For a given site, the neighborhood coincides with that of the regular lattice if no link has been rewired, and this happens with probability $(1-p)^R$, so the ``weight'' of the mean-field contribution should  in principle be $q(p)=1-(1-p)^y$, with $y=R=7$. As shown in Fig.~\ref{fig:126mf}, one obtains a bifurcation diagram more similar to the experimental one (for what concerns the localization of bifurcations) for $y=5.55$. 

\section{Conclusions}~\label{sec:last}
We have obtained the conditions under which a probabilistic, totalistic cellular automaton approximates, in the mean-field limit, the logistic equation. We have shown that the full range of the logistic map parameter $a$ can only by accomplished in the limit of an infinite-range ($R\rightarrow\infty$) neighborhood. 

We have then shown that, exploiting the Watts-Strogatz ``small-world'' mechanism, one can obtain a good approximation of the logistic behavior for a one-dimensional cellular automaton already with a fraction $p$ of rewired links of about 60\%, for various values of the parameter $a$.

For $p=1$, the actual implementation of the cellular automaton exhibits errors between the numerical and the theoretical values of the density that scale as $N^{-1}$. Therefore, the closest approximation of a one-dimensional cellular automaton to the logistic map is  reached for a large (infinite) number of cells, large (infinite) neighborhood size and a sufficiently large (above 60\%) fraction of rewired links.  

We have  also investigated the bifurcation cascade as a function of the fraction rewired links $p$, and shown that this scenario also holds for a deterministic, totalistic CA with the same basic symmetries of the probabilistic one, with either a random choice of neighbors at each time, or a quenched (small-world) connection matrix.

\backmatter

\section*{Declarations}

\begin{itemize}
\item Funding: not applicable
\item Conflict of interest/Competing interests: not applicable
\item Ethics approval and consent to participate: not applicable
\item Consent for publication: not applicable
\item Data availability: not applicable
\item Materials availability: not applicable
\item Code availability: simulation codes (C language) are available upon request
\item Author contribution: only one author
\end{itemize}

\bibliography{ca}

\end{document}